\documentclass[twocolumn,superscriptaddress,PRB]{revtex4}

\usepackage{color}

\usepackage{amsmath}
\usepackage{amssymb}
\usepackage[pdftex]{graphicx}
\usepackage{here}
\usepackage{bm}

\begin{document}

\title{Bipartite Electronic Superstructures in the Vortex Core of Bi$_2$Sr$_2$CaCu$_2$O$_{8+\delta}$}

\author{T. Machida}
\email{tadashi.machida@riken.jp}
\affiliation{RIKEN Center for Emergent Matter Science, Wako, Saitama 351-0198, Japan}

\author{Y. Kohsaka}
\affiliation{RIKEN Center for Emergent Matter Science, Wako, Saitama 351-0198, Japan}

\author{K. Matsuoka}
\affiliation{RIKEN Center for Emergent Matter Science, Wako, Saitama 351-0198, Japan}
\affiliation{Department of Applied Physics, The University of Tokyo, Hongo, Bunkyo-ku, Tokyo 113-8656, Japan}

\author{K. Iwaya}
\affiliation{RIKEN Center for Emergent Matter Science, Wako, Saitama 351-0198, Japan}

\author{T. Hanaguri}
\email{hanaguri@riken.jp}
\affiliation{RIKEN Center for Emergent Matter Science, Wako, Saitama 351-0198, Japan}

\author{T. Tamegai}
\affiliation{Department of Applied Physics, The University of Tokyo, Hongo, Bunkyo-ku, Tokyo 113-8656, Japan}

\begin{abstract}
A magnetic field applied to type-II superconductors introduces quantized vortices that locally quench superconductivity, providing a unique opportunity to investigate electronic orders that may compete with superconductivity.
This is especially true in cuprate superconductors in which mutual relationships among superconductivity, pseudogap, and broken-spatial-symmetry states have attracted much attention.
Here we observe energy and momentum dependent bipartite electronic superstructures in the vortex core of Bi$_2$Sr$_2$CaCu$_2$O$_{8+\delta}$ using spectroscopic-imaging scanning tunneling microscopy (SI-STM).
In the low-energy range where the nodal Bogoliubov quasiparticles are well-defined, we show that the quasiparticle scattering off vortices generates the electronic superstructure known as ``vortex checkerboard''.
In the high-energy region where the pseudogap develops, vortices amplify the broken-spatial-symmetry patterns that preexist in zero field.
These data reveal canonical {\textit d}-wave superconductivity near the node, yet competition between superconductivity and broken-spatial-symmetry states near the antinode.
\end{abstract}

\maketitle

The electronic states in cuprates exhibit distinct features depending on energy and momentum as schematically shown in Fig.~1A~\cite{Keimer2015Nature, Kohsaka2008Nature}.
The low-energy near-nodal states host the homogeneous $d$-wave superconductivity that manifests itself in the Bogoliubov quasiparticle interference (BQPI) patterns imaged by SI-STM~\cite{Hoffman2002Science_QPI, Wang2003PRB, McElroy2003Nature,Kohsaka2008Nature}.
BQPI is no longer observed above the doping-dependent extinction energy $\Delta_0$ and outside the diagonal line in momentum space connecting $(\pi/a_0,0)$ and $(0,\pi/a_0)$, where $a_0$ denotes Cu-O-Cu distance~\cite{Kohsaka2008Nature}.
The higher energy states near the antinode are governed by the pseudogap whose apparent magnitude $\Delta_1$ is spatially inhomogeneous and the quasiparticle excitations near $\Delta_1$ break rotational and translational symmetries of the CuO$_2$ plane~\cite{Kohsaka2008Nature, Lawler2010Nature, Fujita2014PNAS}.

In order to establish the relationship among these electronic states, it is indispensable to investigate how the pseudogap and the broken-spatial-symmetry state are affected when the superconductivity is suppressed.
Introduction of vortices is one of the ways to suppress superconductivity.
It has been shown in La- and Y-based cuprates that the electronic orders that break the spatial symmetry are enhanced or even generated by vortices~\cite{Lake2001Science, Chang2012NatPhys, Blanco-Canosa2013PRL, Gerber2015condmat, Wu2011Nature, Wu2013NatureCommun}.
However, the detailed energy, spatial and momentum-space structures of the vortex-enhanced orders are unknown.
To address this issue, we utilize SI-STM in the vortex state owing to the following three advantages.
First, vortices can suppress superconductivity at the lowest temperatures where the thermal broadening effect is negligible, making it possible to study the precise energy scale of the phenomenon.
Second, atomic-scale spatial resolution of SI-STM is highly beneficial not only to identify the locations of vortices but also to determine the real-space structure of vortex-induced states.
Finally, by using the Fourier transformation, SI-STM acquires the momentum-space resolution even under magnetic fields that enables us to discuss the near-nodal and antinodal states separately.

We choose optimally doped Bi$_2$Sr$_2$CaCu$_2$O$_{8+\delta}$ (superconducting transition temperature $T_c\sim90$~K) as a sample because of its high quality surface necessary for SI-STM~\cite{SM}.
Pioneering SI-STM studies of the vortices in Bi-based cuprates have discovered that an electronic superstructure, so-called ``vortex checkerboard'', is nucleated in the vortex core~\cite{Hoffman2002Science_vortex, Matsuba2007JPSJ, Yoshizawa2013JPSJ}.
Although a possible connection between the ``vortex checkerboard'' and the electronic order that breaks spatial symmetry has been discussed~\cite{Hoffman2002Science_vortex, Matsuba2007JPSJ, Yoshizawa2013JPSJ}, the electronic states in the vortex core are still elusive, probably because the energy ranges so far studied are mostly below $\Delta_0$ and little is known about the momentum-space electronic states in the vortex state.
Here we reveal energy-dependent bipartite nature of the electronic superstructures in the vortex core using the Fourier-transform SI-STM over a wide energy range.

\begin{figure*}[t]
\begin{center}
\includegraphics[width=14cm]{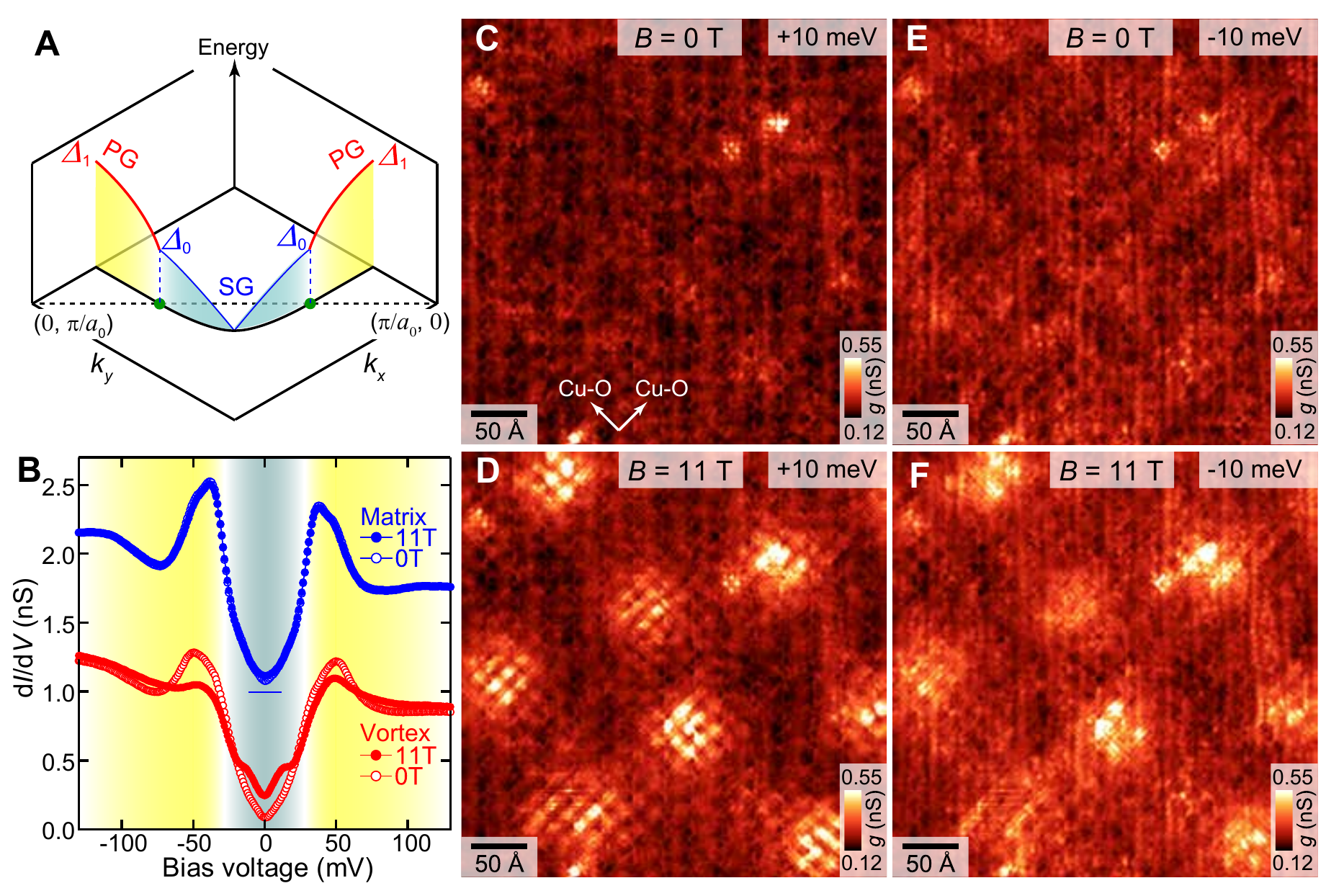}
\end{center}
\caption{
\textbf{Nodal-antinodal dichotomy and spectroscopic features of vortices.}
(\textbf{A}) Schematic illustration of the excitation gap in momentum space showing the dichotomy between the $d$-wave superconductivity near the node (light blue area) and the antinodal states governed by the pseudogap (yellow area).
These two regimes are separated by the line connecting ($\pi/a_0$,0) and (0,$\pi/a_0$). (\textbf{B}) Comparison between tunneling spectra taken at $B=0$~T (open symbols) and 11~T (solid symbols).
The measurement temperature is 4.6~K.
Red and blue data depict the spectra spatially averaged over the regions near vortices and far from vortices, respectively.
Detailed point spectra before averaging are shown in fig.~S3.
(\textbf{C} and \textbf{D})  Differential conductance maps $g({\mathbf r}, E, B)$ at $E=+10$~meV in magnetic fields $B=0$~T and 11~T, respectively.
White arrows in (C) denote the Cu-O boding directions.
Vortices and their internal structures (``vortex checkerboard'') are clearly imaged in (D).
(\textbf{E} and \textbf{F}) Differential conductance maps $g({\mathbf r}, E, B)$ at $E=-10$~meV in magnetic fields $B=0$~T and 11~T, respectively.
The tunneling conductance at each location was obtained by numerical differentiation of the current-voltage characteristics and by post-smoothing with the energy window of $\pm$2~meV.
}
\end{figure*}

Figure~1, C to F show differential tunneling conductance $g({\mathbf r},E,B)$ maps at energy $E=\pm10$~meV taken in magnetic fields of $B=0$~T and 11~T in exactly the same field of view.
Here, ${\mathbf r}$ denotes the position on the surface.
Vortex cores are identified as enhanced $g({\mathbf r},E,B)$ regions with the ``vortex checkerboard'' structure~\cite{Hoffman2002Science_vortex, Matsuba2007JPSJ, Yoshizawa2013JPSJ}.
As shown in Fig.~1B, the vortex alters the spectrum in two different energy regions~\cite{Matsuba2007JPSJ, Yoshizawa2013JPSJ, Renner1998PRL, Pan2000PRL, Matsuba2003JPSJ}: the emergence of conductance humps around $|E|\sim10{\rm~meV}<\Delta_0$ and the suppression of the peaks at $\Delta_1$.
Since these two energy regions occupy different sectors in momentum space, near-nodal and antinodal states (Fig.~1A), it is intriguing to explore the momentum-space characters of the vortex-induced electronic states.

By taking the Fourier transformation from the spectroscopic images, we can estimate the characteristic wavevectors ${\mathbf q}(E,B)$'s of electronic-state modulations.
However, in heterogeneous systems such as Bi$_2$Sr$_2$CaCu$_2$O$_{8+\delta}$, $g({\mathbf r}, E, B)$ not only reflects the ${\mathbf r}$ dependence of the local density-of-states (LDOS) at $E$ but also includes LDOS modulations at different energies because of the ${\mathbf r}$-dependent tip elevation associated with the feed-back loop.
This so-called set-point effect can be suppressed by taking a ratio $Z({\mathbf r},E,B) = g({\mathbf r},+|E|,B)/g({\mathbf r},-|E|,B)$, which faithfully represents the ratio of the LDOS at $\pm |E|$~\cite{Kohsaka2007Science, Hanaguri2007NatPhys}.

First, we focus on the low-energy near-nodal region and argue the origin of the ``vortex checkerboard''.
At $B=0$~T, the only relevant phenomenon near the node is the BQPI that is described by the ``octet model''~\cite{Wang2003PRB, McElroy2003Nature} in which the eight tips of the banana-shaped constant-energy contours in momentum space dominate the quasiparticle scatterings, resulting in a set of energy-dispersive characteristic wavevectors ${\mathbf q}_i$ ($i$ = 1, 2,$\cdots$, 7)(See section 2.1 of supplementary online text).
Figure~2A depicts the typical BQPI pattern seen in $Z_q({\mathbf q},E,B=0{\rm~T})$, the Fourier-transformed image of $Z({\mathbf r},E,B=0{\rm~T})$, showing the octet ${\mathbf q}_i$'s.
(In the optimally doped Bi$_2$Sr$_2$CaCu$_2$O$_{8+\delta}$, signals at ${\mathbf q}_4$ and ${\mathbf q}_5$ are weak in $Z$~\cite{Kohsaka2008Nature,Fujita2014Science}.)
We have performed standard BQPI analysis~\cite{McElroy2003Nature,Kohsaka2008Nature} (See section 2.1 of supplementary online text), and confirm that the BQPI is restricted in the near-nodal region below the extinction energy $\Delta_0\sim30$~meV~\cite{Kohsaka2008Nature}.

\begin{figure}[t]
\begin{center}
\includegraphics[width=8.5cm]{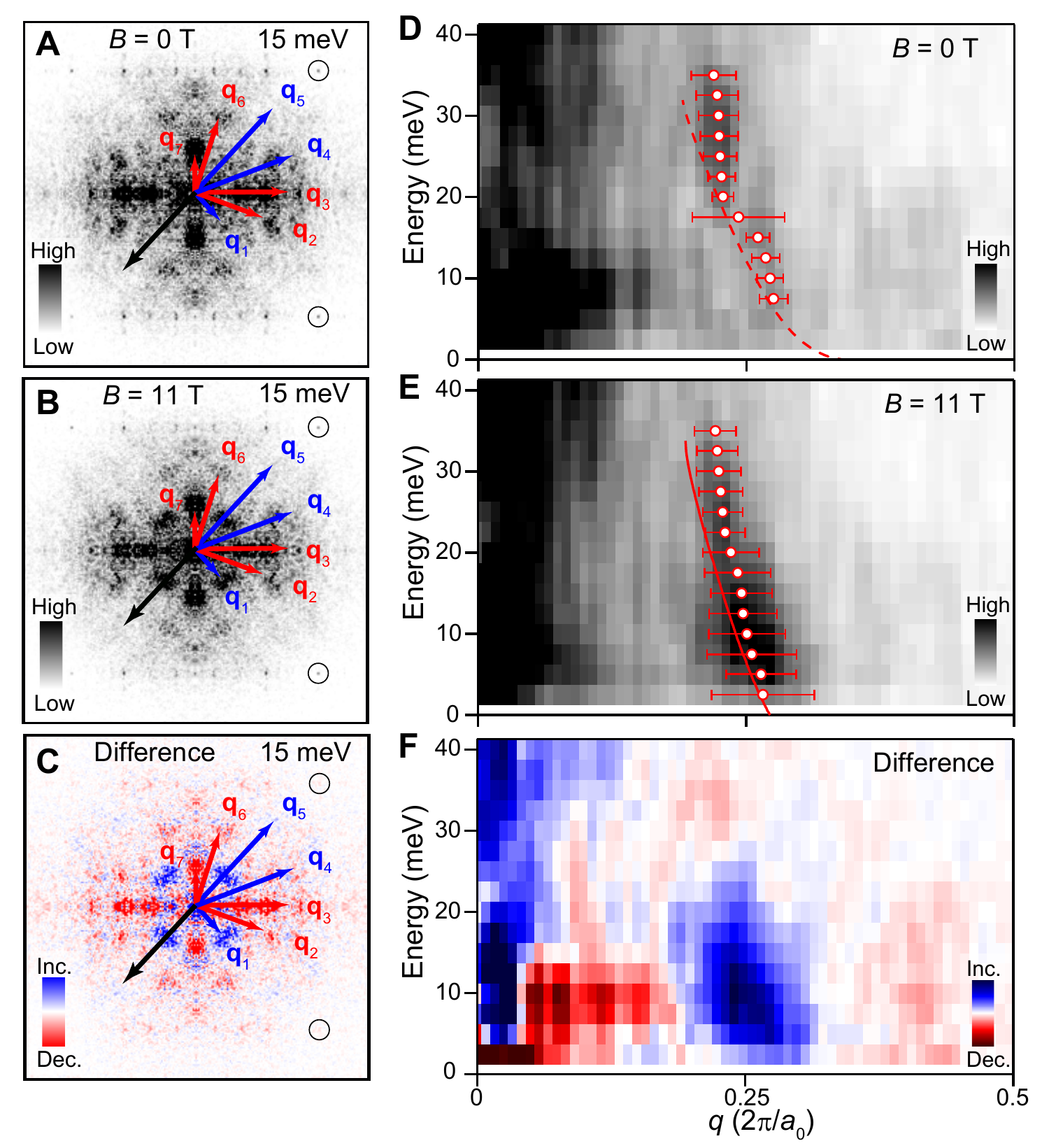}
\end{center}
\caption{
\textbf{Magnetic-field effect on the low-energy state showing the BQPI origin of the ``vortex checkerboard''.}
(\textbf{A} and \textbf{B}) Two-fold symmetrized Fourier-transformed images $Z_q({\mathbf q}, E, B)$ taken in the field of view of 470 $\times$ 470 \AA$^2$.
The tunneling conductance at each location was taken by the standard lock-in technique with the modulation amplitude of 2.5~mV$_{\rm rms}$.
The measurement temperature is 4.6~K.
Red and blue arrows indicate the sign-reversing (${\mathbf q}_2$, ${\mathbf q}_3$, ${\mathbf q}_6$, ${\mathbf q}_7$) and sign-preserving scattering (${\mathbf q}_1$, ${\mathbf q}_4$, ${\mathbf q}_5$) wavevectors, respectively.
Black circles show the Bragg spots.
(\textbf{C}) Difference between (A) and (B); $Z_q({\mathbf q}, E=15{\rm~meV}, B=11{\rm~T})-Z_q({\mathbf q}, E=15{\rm~meV}, B=0{\rm~T})$.
(\textbf{D} and \textbf{E}) Energy-dependent line profiles taken along the black arrows in (A) and (B), respectively.
Red open circles denote the position of ${\mathbf q}_1$ peak that has been determined by fitting the line profile at each energy by Lorentzian function.
Error bars indicate the full-width-half-maximum of the ${\mathbf q}_1$ peak.
Dashed red line in (D) and solid red line in (E) are the dispersions calculated from ${\mathbf q}_2$, ${\mathbf q}_3$, ${\mathbf q}_6$, ${\mathbf q}_7$ based on the ``octet model''.
Note that the vortex-induced signal exhibits the dispersion that is consistent with the BQPI.
(\textbf{F}) Energy-dependent line profile taken along the black arrow in (C), showing the magnetic-field induced change.
See section 2.1 of supplementary online text for details.
}
\end{figure}

An important question here is whether the ``vortex checkerboard'', which is most prominent around $|E|\sim10{\rm~meV}<\Delta_0$, represents an electronic order or not.
To answer this question, we have repeated the same $Z({\mathbf r},E,B)$ analysis at $B=11$~T in the same field of view (See section 2.1 of supplementary online text).
As shown in Fig.~2B, no additional peak is detected whereas the intensity of each ${\mathbf q}_i$ peak depends on the magnetic field.
Figure~2C highlights the field-induced change obtained by subtracting $Z_q({\mathbf q},E,B=0{\rm~T})$ from $Z_q({\mathbf q},E,B=11{\rm~T})$.
The enhanced intensity appears at ${\mathbf q}_1$, which represents the wavevector of the ``vortex checkerboard''.
We note that the intensities at ${\mathbf q}_2$, ${\mathbf q}_3$, ${\mathbf q}_6$, and ${\mathbf q}_7$ are suppressed by the magnetic field.
All of these scattering ${\mathbf q}_i$'s reverse the sign of the $d$-wave superconducting gap between the initial state and the final state, while the sign is preserved in the the case of ${\mathbf q}_1$ scattering.
Such suppression and enhancement of sign-reversing and sign-preserving scatterings, respectively, are exactly what are expected from the coherence factors of the quasiparticle scatterings off vortices~\cite{Hanaguri2009Science, Pereg-Barnea2008PRB, Maltseva2009PRB}, suggesting that the vortex-enhanced BQPI is the cause of the ``vortex checkerboard''.

The BQPI scenario is further supported by the energy dispersion in the vortex-enhanced signal at ${\mathbf q}_1$ (Fig.~2, D and E).
The observed ${\mathbf q}_1$ dispersion agrees well with the behavior calculated from ${\mathbf q}_2$, ${\mathbf q}_3$, ${\mathbf q}_6$, and ${\mathbf q}_7$ based on the octet model.
Together with the fact that the vortex-enhanced signal diminishes near $\Delta_0$ (Fig.~2F), we ascribe the ``vortex checkerboard'' to the BQPI and conclude that no electronic order is nucleated in the vortex cores  at $E<\Delta_0$.
We note that other Friedel-type oscillations such as bound-state oscillations in the quantum-limit vortex core~\cite{Yoshizawa2013JPSJ} may also be relevant (See section 2.2 of supplementary online text).

Next, we study the effect of vortices near $\Delta_1$.
The electronic feature here is characterized by the bond-centered unidirectional electronic entity that breaks both rotational and translational symmetries~\cite{Kohsaka2008Nature, Lawler2010Nature, Fujita2014PNAS, Kohsaka2007Science}, being reminiscent of the short-range charge order detected by X-ray scattering~\cite{Comin2014Science, daSilvaNeto2014Science}.
For the purpose of brevity, we call this electronic entity as ``nanostripe'' hereafter.
To visualize the ``nanostripe'' which is most prominent at $\Delta_1$, we follow the procedure used in \cite{Kohsaka2008Nature,Lawler2010Nature}.
Since $\Delta_1$ is spatially inhomogeneous, we first normalize $E$ by local $\Delta_1({\mathbf r})$ and map $Z({\mathbf r},e \equiv E/\Delta_1({\mathbf r})=1,B)$.
Figure~3A shows $Z({\mathbf r},e=1,B=0{\rm~T})$ in the same field of view as Fig.~1.
The ``nanostripe'' in the optimally-doped sample is weak in intensity~\cite{Fujita2014Science} and is observed only in the limited regions where $Z({\mathbf r},e=1,B)$ is larger.
We find that these regions have larger $\Delta_1({\mathbf r})$ and vortices tend to reside there, suggesting a vortex pinning mechanism associated with the pseudogap (See section 2.3 of supplementary online text).
We have repeated the same measurement at $B=11$~T (Fig.~3B) and have revealed that $Z({\mathbf r},e=1,B)$ is enhanced in the vortex cores (Fig.~3C).
As shown in the insets of Fig.~3, A to C, the structure of the ``nanostripe'' that preexists in the absence of the vortex core is unchanged but its contrast is enhanced.

\begin{figure*}[t]
\begin{center}
\includegraphics[width=14cm]{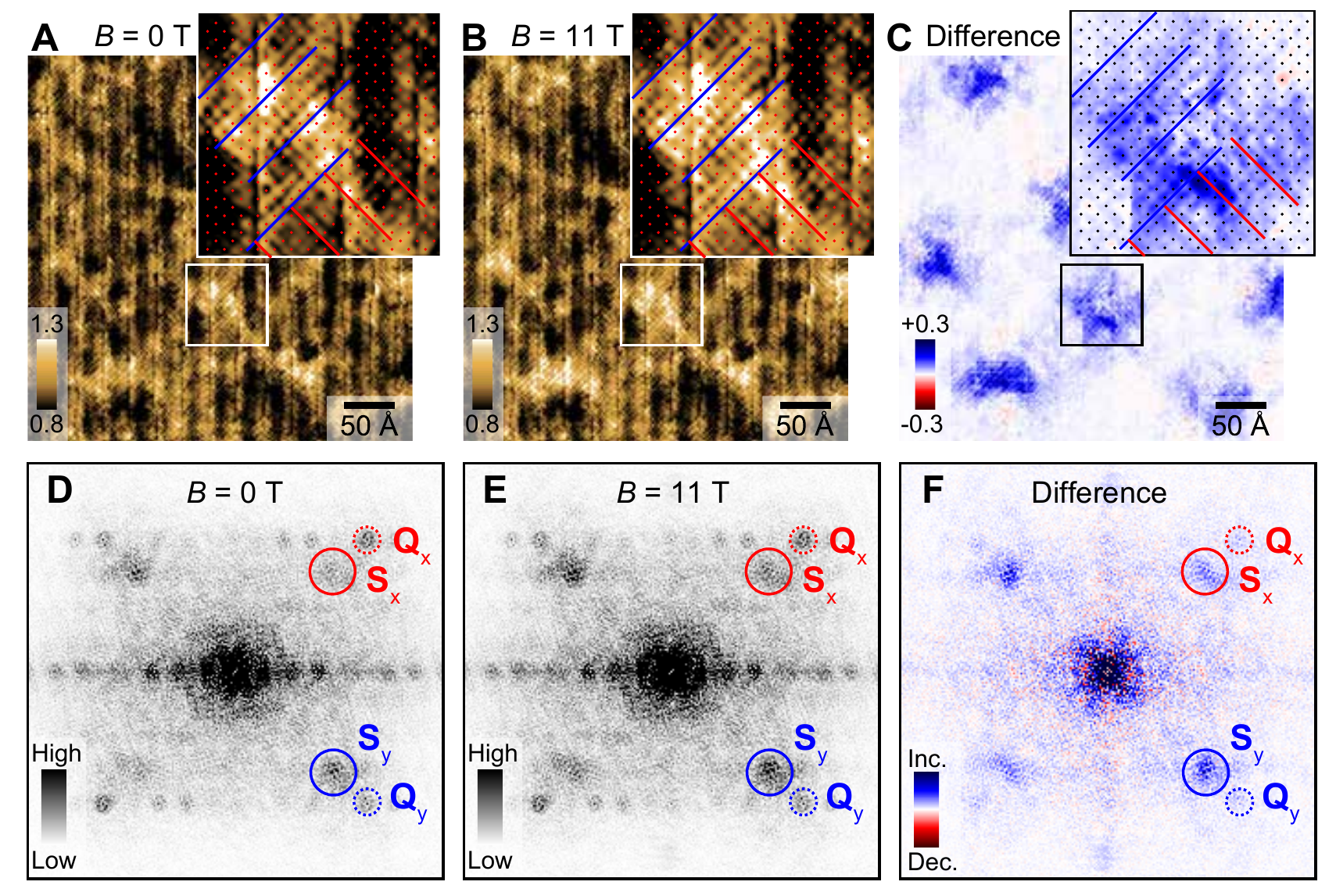}
\end{center}
\caption{
\textbf{Magnetic-field effect on the ``nanostripe'' at the pseudogap energy scale.}
(\textbf{A} and \textbf{B}) Conductance ratio maps $Z({\mathbf r}, e=1, B)$ in magnetic fields $B=0$~T and 11~T, respectively.
These images are taken in the same field of view as Fig.~1, (C) to (F).
The tunneling conductance at each location was obtained by numerical differentiation of the current-voltage characteristics and by post-smoothing with the energy window of $\pm$10 meV.
The measurement temperature is 4.6~K.
(\textbf{C}) Difference between (A) and (B); $Z({\mathbf r}, e=1, B=11{\rm~T})-Z({\mathbf r}, e=1, B=0{\rm~T})$.
Insets of (A) to (C) are the magnified images of the regions marked by boxes in the main figures.
Red and blue lines in the insets indicate the directions of the ``nanostripe''.
Dots denote the locations of the Cu atoms.
(\textbf{D} to \textbf{E}) Fourier-transformed images $Z_q({\mathbf q}, e=1, B)$ in magnetic fields $B=0$~T and 11~T, respectively.
The field of views are restricted in the vicinity of the vortices by applying a mask~\cite{SM}.
(\textbf{F}) Difference between (D) and (E).
In these Fourier-transformed images, dotted and solid circles indicate the locations of ${\mathbf Q}_{x,y}=(2\pi/a_0, 0), (0, 2\pi/a_0)$ and ${\mathbf S}_{x,y}\sim3/4\times(2\pi/a_0, 0), 3/4\times(0, 2\pi/a_0)$, respectively.
}
\end{figure*}

It has been known that the ``nanostripe'' consists of two sets of wavevectors~\cite{Lawler2010Nature}: ${\mathbf Q}_{x,y}=(2\pi/a_0,0), (0,2\pi/a_0)$ whose inequivalent intensities represents the degree of broken rotational symmetry and ${\mathbf S}_{x,y}\sim(3/4\times(2\pi/a_0),0), (0,3/4\times(2\pi/a_0))$ that features the broken translational symmetry.
To test which broken symmetry is affected in the vortex core, we perform the Fourier analysis.
By applying a mask generated from the image of vortices, we restrict our field of view in the vicinity of the vortex core to effectively extract the vortex-enhanced features~\cite{SM}.
As shown in Fig.~3, D and E, Fourier peaks corresponding to ${\mathbf Q}$ and ${\mathbf S}$ are identified at both $B=0$~T and 11~T.
In the difference image shown in Fig.~3F, the intensity at ${\mathbf S}$ is enhanced, whereas the change at ${\mathbf Q}$ is small.
This observation indicates that the vortex core predominantly amplifies the translational-symmetry breaking.

The bipartite vortex-induced changes observed at $E<\Delta_0$ and at $E\sim\Delta_1$ provide important implications to understand the electronic-state competition in cuprates.
First, the BQPI origin of the ``vortex checkerboard'' at $E<\Delta_0$ suggests that canonical phenomenology of $d$-wave superconductivity applies in the near-nodal region even when the vortices are introduced.

At $E\sim\Delta_1$ near the antinode, the ``nanostripe'' amplified in the vortex core resembles the magnetic-field-enhanced charge orders in Y-based cuprates~\cite{Chang2012NatPhys, Blanco-Canosa2013PRL, Gerber2015condmat, Wu2011Nature, Wu2013NatureCommun}.
This indicates that the magnetic-field-enhanced broken-spatial-symmetry state is universal in cuprates.

It is noteworthy that even though the contrast of the ``nanostripe'' seen in $Z({\mathbf r},e=1,B)$ is enhanced near the vortex core, the spectral weight at $|E|\sim\Delta_1$ is suppressed as shown in Fig.~1B.
If the missing weight is caused by the reduction of electrons condensed in the pseudogap state, the enhanced ``nanostripe'' in the vortex core should be distinct from the pseudogap formation.
This picture is unlikely because the previous SI-STM results in the absence of vortices show close correlation between the pseudogap and the ``nanostripe''~\cite{Kohsaka2008Nature, Fujita2014Science, Kohsaka2012NatPhys}.
Rather, we infer that Bogoliubov quasiparticles exist even near the antinode at $|E|\sim\Delta_1$ and that the the missing weight is originated from the suppression of superconductivity~\cite{Yoshizawa2013JPSJ,He2014Science}.
Since the pseudogap or the ``nanostripe'' also develops at the same energy scale and two different orders do not share the same region in energy-momentum space, it is plausible that superconductivity and the pseudogap are nearly degenerated but competing states near the antinode~\cite{Chang2012NatPhys,Wu2013NatureCommun,He2011Science,He2014Science}.
This scenario is consistent with the heterogeneous electronic states in real space because randomly-dispersed dopants and defects locally perturb the balance between superconductivity and the pseudogap, and consequently bring about the nano-scale mixture of them.
The observed vortex-enhanced ``nanostripe'' suggests that the vortex core is one of these perturbations that affects the competition near the antinode.
Since the energy scales of superconductivity and the pseudogap near the antinode change with doping, we anticipate that the vortex-core electronic states at different doping levels provide more insights into the relationship among superconductivity, pseudogap, and the ``nanostripe'', leading to establishing the phenomenology of cuprates.




\section*{Acknowledgments}
The authors thank J. C. Davis, A. N. Pasupathy, and S. Tajima for discussions and comments.
This work was partly supported by Grant-in-Aid for Scientific Research from Japan Society for the Promotion of Science (Grant No. 20244060).

  \makeatletter
    \renewcommand{\theequation}{S\arabic{equation}}
  \makeatother

  \makeatletter
    \renewcommand{\thefigure}{S\arabic{figure}}
  \makeatother

\section*{Supplementary Materials}

\section{Materials and Methods}

\subsection{Procedure of SI-STM experiments}

SI-STM experiments were performed at a temperature of 4.6~K with a commercial low-temperature ultra-high-vacuum STM (Unisoku USM-1300) modified by ourselves~\cite{Hanaguri2006JPhys}.
Single crystals of Bi$_2$Sr$_2$CaCu$_2$O$_{8+\delta}$ were grown by the traveling-solvent floating zone method and were annealed to have optimal hole concentration.
The superconducting transition temperature $T_c\sim90$~K was determined by the magnetization measurement.
The sample was cleaved \textit{in-situ} at $\sim$77~K to obtain a clean and flat (001) surface and was transferred quickly to the STM unit kept at 4.6~K.
We used an electro-chemically etched tungsten wire as an STM tip, which was cleaned and characterized \textit{in-situ} with a field-ion microscope.
All the SI-STM data were taken with the feed-back set point at a sample bias voltage of -150~mV and a tunneling current of 150~pA.
Magnetic field was applied perpendicular to the cleaved (001) surface.
Whenever we changed the magnetic field, the sample was heated up to $\sim$30~K to make the vortex distribution inside the sample uniform.

\subsection{Procedure to obtain $\bm{Z({\mathbf r},E,B)}$ and $\bm{Z({\mathbf r},e,B)}$}
It is well-known that electronic states of Bi$_2$Sr$_2$CaCu$_2$O$_{8+\delta}$ are heterogeneous especially near $\Delta_1$.
Therefore, in order to argue the vortex-induced local changes in the electronic state, it is indispensable to compare two SI-STM data sets, with and without a magnetic field, in exactly the same field of view.
This is challenging because the actual SI-STM images are inevitably distorted in an uncontrollable manner due to the creeping of the piezoelectric scanner, etc.
In order to obtain distortion-free images from the observed ones, we utilize the so-called Lawler-Fujita algorithm~\cite{Lawler2010Nature}, which estimates the local distortion as a phase shift in the crystal-lattice modulations.
We first estimate and correct the distortions in the topographic images simultaneously taken with the $g({\mathbf r},E,B)$ images and correct the spectroscopic images, $g({\mathbf r},E,B)$ and $Z({\mathbf r},E,B)$, using the same local distortions.
Various images obtained by SI-STM after the correction are shown in Fig.~S1.

In order to obtain $Z({\mathbf r},e,B)$, we have to estimate $\Delta_1(\mathbf{r})$ at each pixel.
We define $\Delta_1(\mathbf{r})$ at the energy of the peak in the unoccupied state in the $g(\mathbf{r},E,B=0{\rm~T})$ spectrum and normalize the energy as $e=E/\Delta_1(\mathbf{r})$.
If there are multiple peaks in the spectrum, we have chosen the highest peak.
We have used $\Delta_1(\mathbf{r})$ estimated at $B = 0$~T to normalize the energy in a field as well.
Even if we use $\Delta_1(\mathbf{r})$ estimated at $B = 11$~T to calculate $Z({\mathbf r},e,B = 11~\mathrm{T})$, the conclusion in this paper is unaffected.

\subsection{Procedure to generate masks}
Vortices are clearly seen in the difference map $\delta g({\mathbf r},E=+10{\rm~meV},B=11{\rm~T})\equiv g({\mathbf r},E=+10{\rm~meV},B=11{\rm~T})-g({\mathbf r},E=+10{\rm~meV},B=0{\rm~T})$ (Fig.~S2A).
We first apply low-pass filter to the $\delta g({\mathbf r},E=+10{\rm~meV},B=11{\rm~T})$ map with a cut-off wavelength of $q\sim 0.05\times 2\pi/a_0$.
Constant contours of this filtered image are used for the boundaries of the masks (Fig.~S2B).

The `vortex' and `matrix' regions used to examine the vortex-induced change are shown in Fig.~S2C.
The spectra shown in Fig.~1B of the main text are the spectra averaged in these regions.
In Fig.~S3, we depict the detailed point spectra near the vortex before averaging.

The mask used for the Fourier analyses (Fig.~3, D to F in the main text) and the restricted $Z({\mathbf r},E,B)$ maps at $B=0$~T and 11~T are indicated in Fig.~S2D, E, and F, respectively.

\section{Supplementary Text}

\subsection{BQPI analyses based on the ``octet model''}
BQPI brings about LDOS modulations which would be reflected on $g({\mathbf r},E,B)$.
However, in actual fact $g({\mathbf r},E,B)$ may also contain extrinsic modulations caused by the set-point effect~\cite{Kohsaka2007Science,Hanaguri2007NatPhys}.
The set-point effect can be largely suppressed in $Z({\mathbf r},E,B)$, which we utilize throughout this work.
The analysis of $Z({\mathbf r},E,B)$ has another advantage in that $Z({\mathbf r},E,B)$ picks up the particular modulations in which $\mathbf{q}(+|E|)=\mathbf{q}(-|E|)$ and the phase difference between the $\mathbf{q}(+|E|)$ and $\mathbf{q}(-|E|)$ modulations is large ($\sim \pi$).
These features are exactly what are expected in the BQPI~\cite{Hanaguri2007NatPhys,Fujita2008PRB}.

We use the ``octet model'' to analyze the BQPI seen in $Z_q({\mathbf q},E,B)$; the Fourier transformed $Z({\mathbf r},E,B)$ ~\cite{Hoffman2002Science_QPI,Wang2003PRB,McElroy2003Nature}.
Here, the seven scattering vectors ${\mathbf q}_i$ ($i$ = 1, 2,$\cdots$, 7) connecting the eight tips of the banana-shaped constant contours in momentum space govern the BQPI because the joint density of states takes maximum for these wavevectors (Fig.~S4A).
As shown in Fig.~2, A and B of the main text and in Fig.~S5, a set of energy-dispersive wavevectors are detected in $Z_q({\mathbf q},E,B)$ and each of the wavevectors can be assigned to one of the octet wavevectors ${\mathbf q}_i$.
The intensities at \textbf{q}$_4$ and \textbf{q}$_5$ are weak, being consistent with the previous reports~\cite{Kohsaka2008Nature,Fujita2014Science}.

To determine the precise locations of ${\mathbf q}_i$'s at $B=0$~T and 11~T, we fit the peaks in $Z_q({\mathbf q},E,B)$ with the two-dimensional Lorentzian plus linear background;

\begin{equation}
f(q_x, q_y) = f_0 + \frac{A}{\left(\frac{q_x - q_{x0}}{\sigma_{x}} \right)^2 + \left(\frac{q_y - q_{y0}}{\sigma_{y}} \right)^2 + 1} + c_x q_x + c_y q_y,
\label{EqS1}
\end{equation}

\noindent
where $f_0$, $c_x$, and $c_y$ are the fitting parameters associated with the linear background, whereas $A$, $q_{x0}$, $q_{y0}$, $\sigma_x$, and $\sigma_y$ are the fitting parameters correspond to the amplitude, the $q_x$ and $q_y$ components of the peak location, and the $q_x$ and $q_y$ components of the half-width-at-half-maximum of the peak, respectively.
Figure~S4B shows the energy dependence of the absolute value of the observed  ${\mathbf q}_i$'s.
The signals at ${\mathbf q}_2$, ${\mathbf q}_3$, ${\mathbf q}_6$, and ${\mathbf q}_7$ diminishes at about 30~meV, which set the extinction energy $\Delta_0$ (Fig.~S4, B and D).
At $E>\Delta_0$, signals at ${\mathbf q}_1$ and ${\mathbf q}_5$ are still there but lose their energy dependence.
The signal near ${\mathbf q}_5$ turns into $\mathbf{S}$, which is one of the ingredients of the ``nanostripe''.

We obtain the normal-state Fermi surface and the superconducting gap dispersion using pairs of (${\mathbf q}_2$, ${\mathbf q}_6$) and  (${\mathbf q}_3$, ${\mathbf q}_7$).
Figure.~S4C depicts the Fermi-surface loci that sustain coherent Bogoliubov quasiparticles.
They are limited inside the diagonal line connecting ($\pi/a_0$,0) and (0, $\pi/a_0$)~\cite{Kohsaka2008Nature} and are hardly affected by a magnetic field.
By contrast a magnetic field suppresses the near-nodal superconducting gap as shown in Fig.~S4D.
These features are consistent with the behaviors observed in the different cuprate superconductor Ca$_{2-x}$Na$_x$CuO$_2$Cl$_2$~\cite{Hanaguri2009Science} and can be associated with the Volovik effect~\cite{Volovik1993JETP}.

In order to check the validity of the ``octet model'' especially in a magnetic field, we have performed the following analyses.
First, we fit the Fermi-surface loci to the quarter circle.
Next, the superconducting gap dispersions $\Delta(\theta)$ are fitted by the following $d$-wave form with an higher-order term.

\begin{equation}
\Delta(\theta)=\Delta_{\rm{BQPI}}[A\cos(2\theta)+(1-A)\cos(6\theta)].
\end{equation}

Here, $\theta$ represents the Fermi-surface angle defined in Fig.~S4C and $\Delta_{\rm{BQPI}}$ and $A$ are fitting parameters.
Using the obtained fitting parameters, the energy dispersions of all the ${\mathbf q}_i$'s can be calculated and plotted in Fig.~S4B and in Fig. ~2, D and E of the main text.
The ``vortex checkerboard'' is characterized by the field-enhanced signal at ${\mathbf q}_1$.
We note that the energy dispersion of ${\mathbf q}_1$ at $B=11$~T well coincides with the calculated one, as in the case at $B=0$~T.
Together with the fact that the ${\mathbf q}_4$ and ${\mathbf q}_5$ are also consistent with the calculated dispersions, we conclude that the field-induced change in the electronic state can be explained in the framework of the ``octet model'' as long as $E<\Delta_0$.

\subsection{Possible origins of the enhanced $\mathbf{q}_1$ modulations}
The applicability of the ``octet model'' implies that the observed electronic-state modulations are associated with the Bogoliubov quasiparticles that reside on the near-nodal Fermi surface.
There is more than one origin to cause such Friedel-type oscillations.
Most naively, the enhanced quasiparticle scattering off vortices may result in the enhanced ${\mathbf q}_1$~\cite{Hanaguri2009Science}.
Another possible origin is the spatial oscillations of the vortex bound state in the quantum-limit vortex core~\cite{Hayashi1998PRL}, as proposed by Yoshizawa and coworkers~\cite{Yoshizawa2013JPSJ}.
This model naturally explains the peaks in the spectrum at $\sim\pm 10$~meV as the discrete bound states.
Nevertheless, more studies are necessary to verify the validity of the bound-state scenario, because it is not clear whether the vortex of a cuprate is in the quantum limit and the formation of the vortex bound state in a $d$-wave superconductor is still a controversial issue~\cite{Franz1998PRL,Kato2002PTP}.

\subsection{Vortex pinning and the competition between the ``nanostripe'' and superconductivity}
It is interesting to examine the relationship between the locations of vortices and the various electronic heterogeneities, since this comparison may give us a hint to identify the elementary process of the vortex pinning.
Although previous SI-STM studies have suggested that vortices in Bi$_2$Sr$_2$CaCu$_2$O$_{8+\delta}$ tend to be pinned at the regions where the $\Delta_1(\mathbf{r})$ is large~\cite{Fukuo2006PRB,Yoshizawa2013JPSJ}, these experiments were conducted only in a field and thereby could not exclude the possibility that the the pseudogap itself would be influenced by vortices.
Here we compare the locations of vortices with the zero-field electronic heterogeneity at $\Delta_1(\mathbf{r})$.

Using the same procedure described in supplementary section~1.3, we first identify the locations of vortices (Fig.~S6, A and B).
We examine the azimuthally averaged cross-correlation function (Fig.~S6C) between the smoothed vortex map Figs~S6B and three different spectroscopic images at $\Delta_1(\mathbf{r})$: spatial variation of $\Delta_1({\mathbf r})$ itself (Fig.~S6D), $Z({\mathbf r},e=1,B)$ (Fig.~S6E), and the local amplitude of the broken-translational-symmetry state of the ``nanostripe'', which is nothing but the local amplitude of the modulations at $S_{x,y}$ (Fig.~S6F).
Here, we define the local amplitude of these modulations $A_{\mathrm{S}}(\mathbf{r})$ as follows,

\begin{eqnarray}
A_{\mathrm{S}}(\mathbf{r}) &\equiv& A(\mathbf{S}_x, \mathbf{r}) + A(\mathbf{S}_y, \mathbf{r}),\nonumber\\
A(\mathbf{S}_\nu, \mathbf{r}) &\equiv& \sum_{\mathbf{r'}}Z(\mathbf{r'})e^{i\mathbf{S}_\nu \cdot \mathbf{r}}f_{\Lambda}(\mathbf{r'}, -\mathbf{r}) \nonumber\\
&\approx& \frac{1}{\sqrt{N}}\sum_{\mathbf{k}}\tilde{Z}(\mathbf{S}_\nu - \mathbf{k})e^{i\mathbf{k} \cdot \mathbf{r}}e^{-\mathbf{k}^2/2\Lambda^2},\ \ (\nu = x\ \mathrm{or}\ y),\nonumber\\
\label{EqS3}
\end{eqnarray}

\noindent
where $f_{\Lambda}(\mathbf{r'} = (\Lambda^2/2\pi)e^{-\Lambda^2|\mathbf{r}|^2/2}$, $1/\Lambda$ is the cut-off length scale, and $\tilde{Z}$ is the complex Fourier transform of $Z({\mathbf r},e = 1,B)$.
As shown in Fig.~S6C, all of these quantities exhibit strong correlations with the locations of vortices.
Since vortices are generally pinned at weakly superconducting regions, these results suggest that at least one of these quantities would represent the fundamental measure of the weakness of superconductivity.
Although the microscopic mechanism of vortex pinning is unclear at present, the observed correlation between the vortex location and $A_{\mathrm{S}}(\mathbf{r})$ clearly indicates that the superconductivity is weak in the region where the ``nanostripe'' is prominent, indicating the competition between superconductivity and the ``nanostripe'' in Bi$_2$Sr$_2$CaCu$_2$O$_{8+\delta}$.

\onecolumngrid
\newpage

\section{Supplementary Figures}

\setcounter{figure}{0}

\begin{figure*}[h]
\begin{center}
\includegraphics[width=16cm]{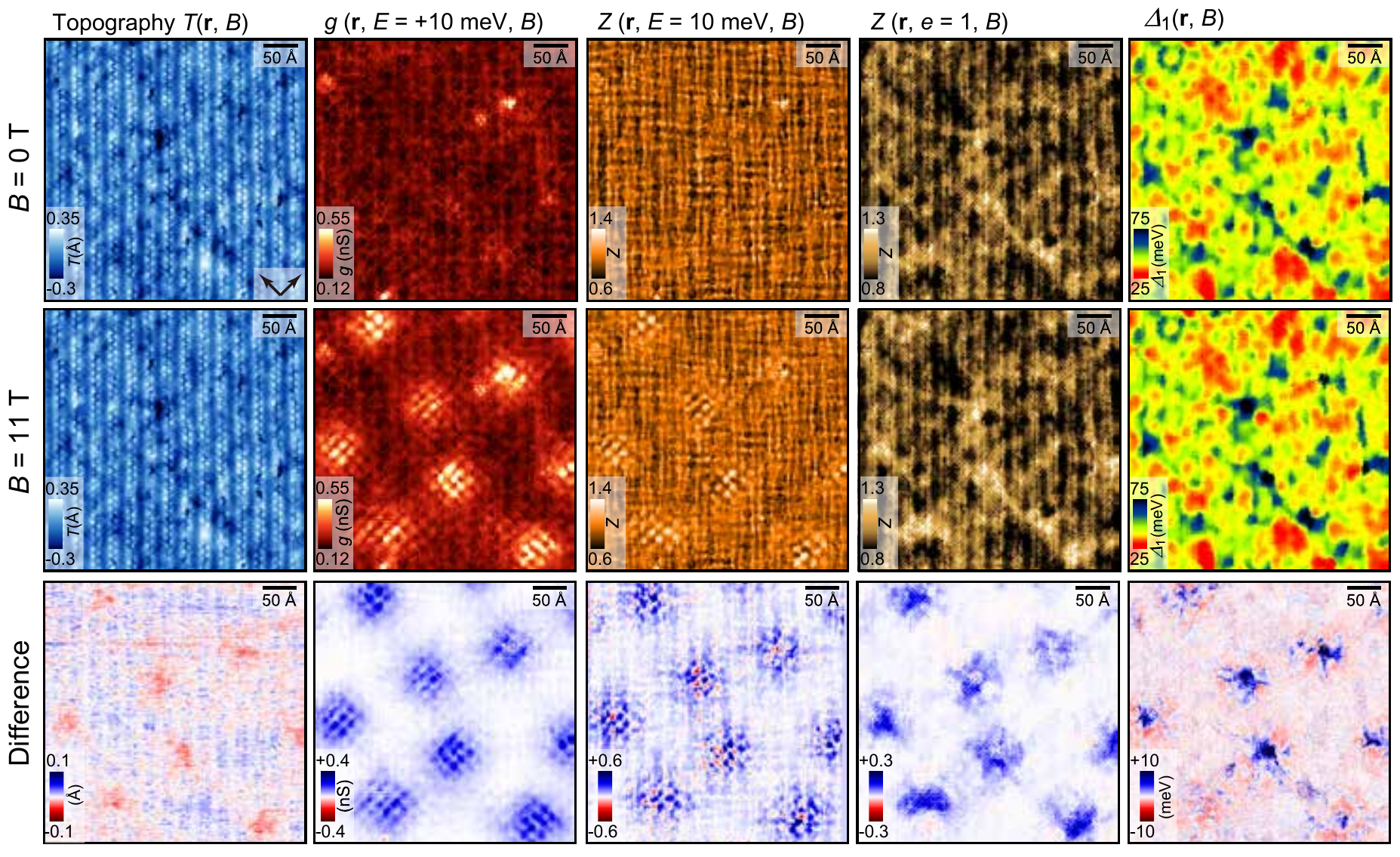}
\end{center}
\caption{
Top and middle rows display two series of SI-STM results on a 380 $\times$ 380 \AA$^2$ field of view at $B= 0$~T and 11~T, respectively.
Bottom row indicates difference images, which are obtained by subtracting the results at 0~T from those at 11~T.
First column; topographic images, second; $g(\mathbf{r},E=+10{\rm~meV},B)$, third; $Z(\mathbf{r},E=10{\rm~meV},B)$, forth; $Z(\mathbf{r},e=1,B)$, and fifth; $\Delta_1({\mathbf r})$ .
Although an apparent enhancement of $\Delta_1(\mathbf{r})$ is observed in some of the vortex cores, changes in the original $g(\mathbf{r},E,B)$ spectra near $\Delta_1({\mathbf r})$ are very small; due to small $dg(\mathbf{r},E,B)/dE$ near $\Delta_1({\mathbf r})$ , small change in $g(\mathbf{r},E,B)$ gives rise to large effect on the estimation of $\Delta_1({\mathbf r})$ .
See, for example, middle spectra along the line C8 in Fig.~S3.
We have used $\Delta_1(\mathbf{r})$ at $B=0$~T to create both of $Z(\mathbf{r},e=1,B=0{\rm~T})$ and $Z(\mathbf{r},e=1,B=11{\rm~T})$ maps.
Black arrows in the topographic image at $B=0$~T denote the Cu-O bonding directions.
}
\label{FigS1}
\end{figure*}

\begin{figure*}[h]
\begin{center}
\includegraphics[width=14cm]{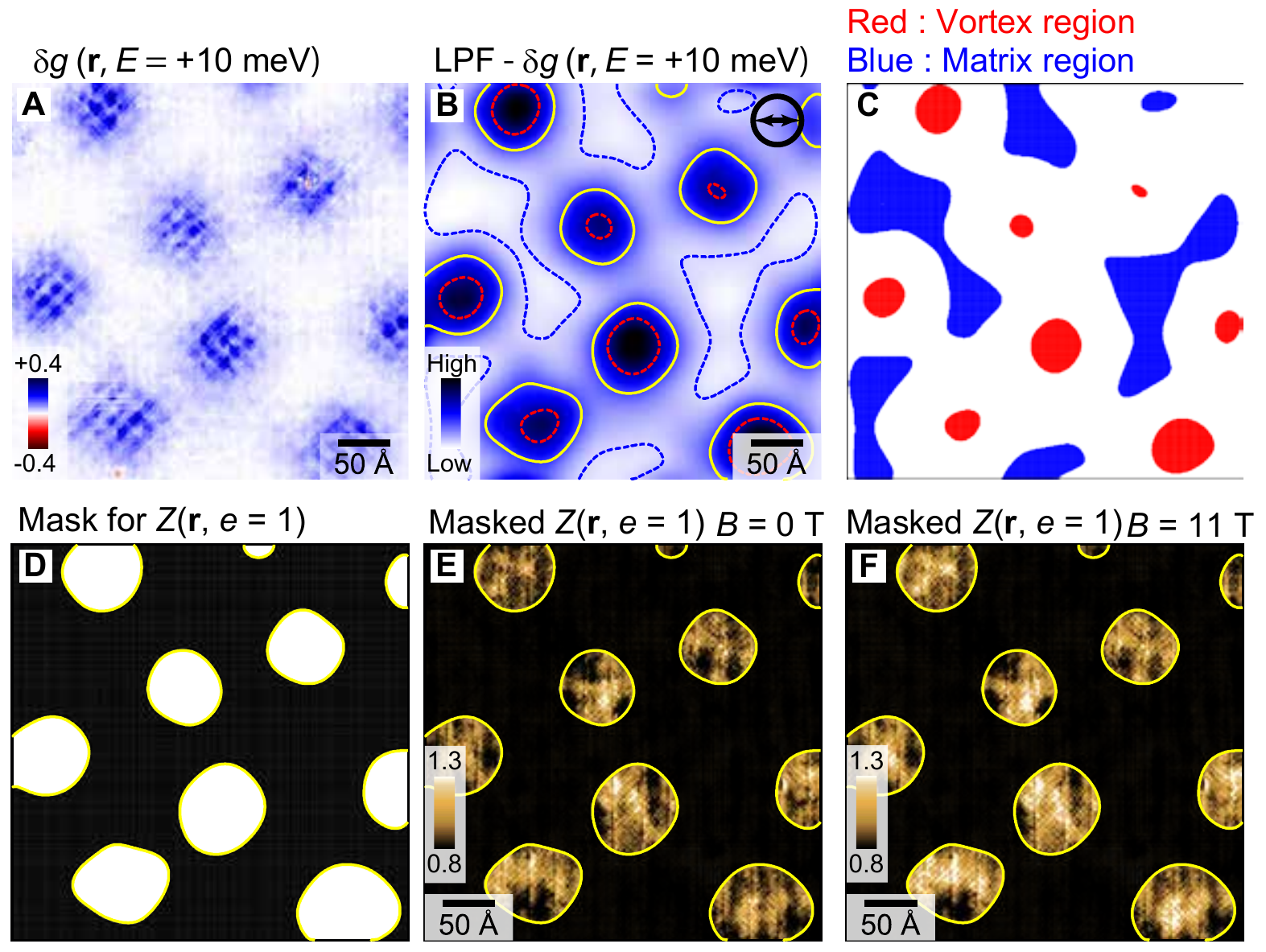}
\end{center}
\caption{
(\textbf{A}) Difference between the conductance maps at $B= 0$~T and 11~T; $\delta g(\mathbf{r},E=+10{\rm~meV}) = g(\mathbf{r},E=+10{\rm~meV}, B=11{\rm~T})-g(\mathbf{r},E=+10{\rm~meV}, B=0{\rm~T})$.
(\textbf{B}), Low-pass filtered image of (A).
A cut-off wavevector is chosen to be $0.05\times 2\pi/a_0$ in the filtering process.
Corresponding spatial resolution is indicated as a black circle and an arrow.
Red dashed, blue dashed, and yellow lines are the contours with the value of 65, 8, 35\% of the difference between the maximum and minimum values of this image, respectively.
(\textbf{C}) Separation of ``vortex'' and ``matrix'' regions for the spatially averaging process used in Fig.~1B of the main text.
Red and blue regions denote ``vortex'' and ``matrix'' regions and are defined as the regions surrounded by the red and blue dashed lines in (B).
(\textbf{D}) A mask for the effective extraction of the field-effect on the ``nanostripe''.
White and black regions denote the ``vortex'' and ``matrix'' regions, which are separated by the yellow contours in (B).
(\textbf{E}) and (\textbf{F}) Masked conductance ratio maps at PG energy at $B=0$~T and 11~T, respectively.
Fourier transformed images of these masked maps are shown in Fig.~3, D and E in the main text, respectively.
}
\label{FigS2}
\end{figure*}

\begin{figure*}[h]
\begin{center}
\includegraphics[width=11cm]{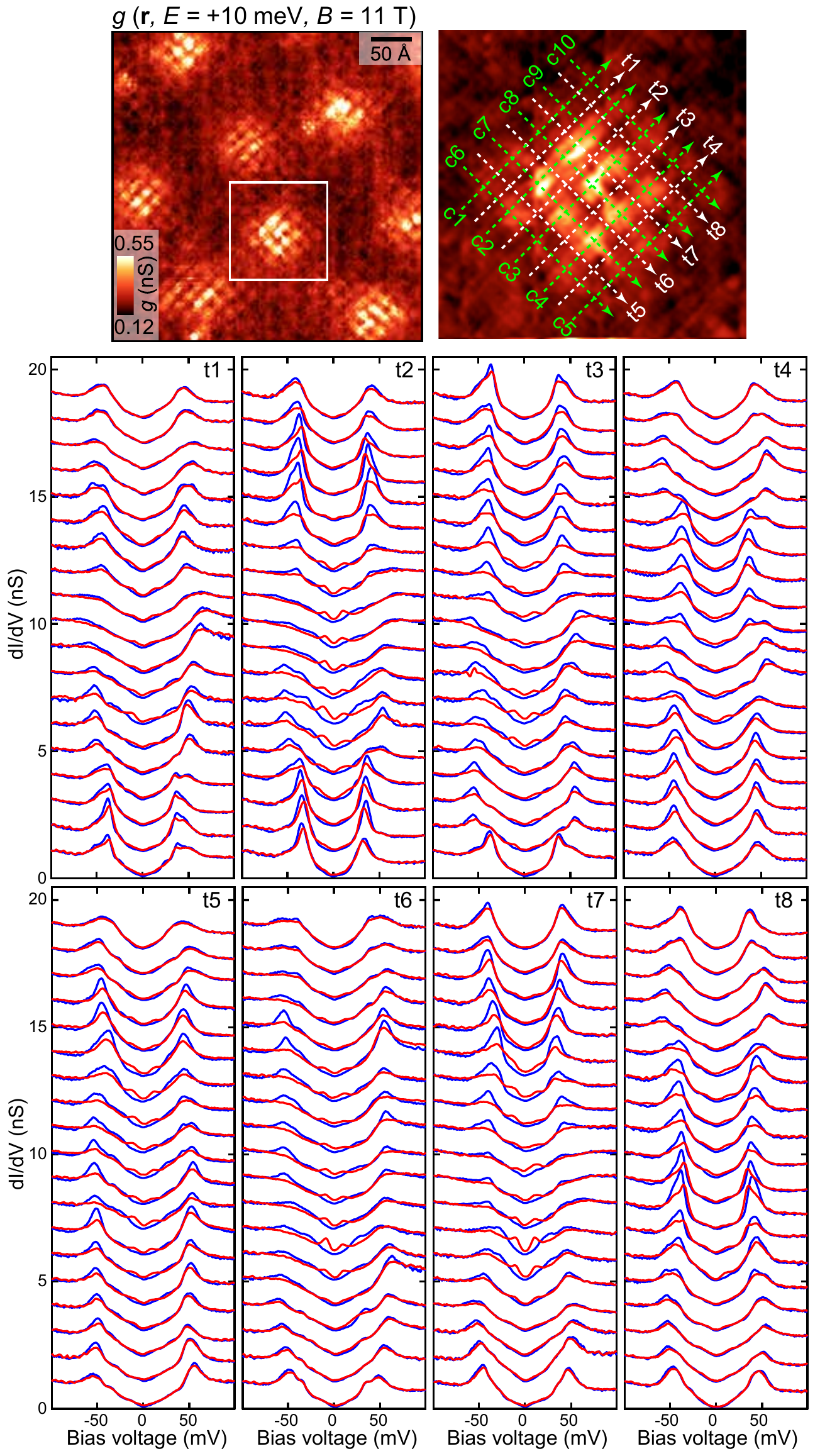}
\end{center}
\caption{
Continued on next page.}
\label{FigS3}
\end{figure*}

\setcounter{figure}{2}
\begin{figure*}[h]
\begin{center}
\includegraphics[width=14cm]{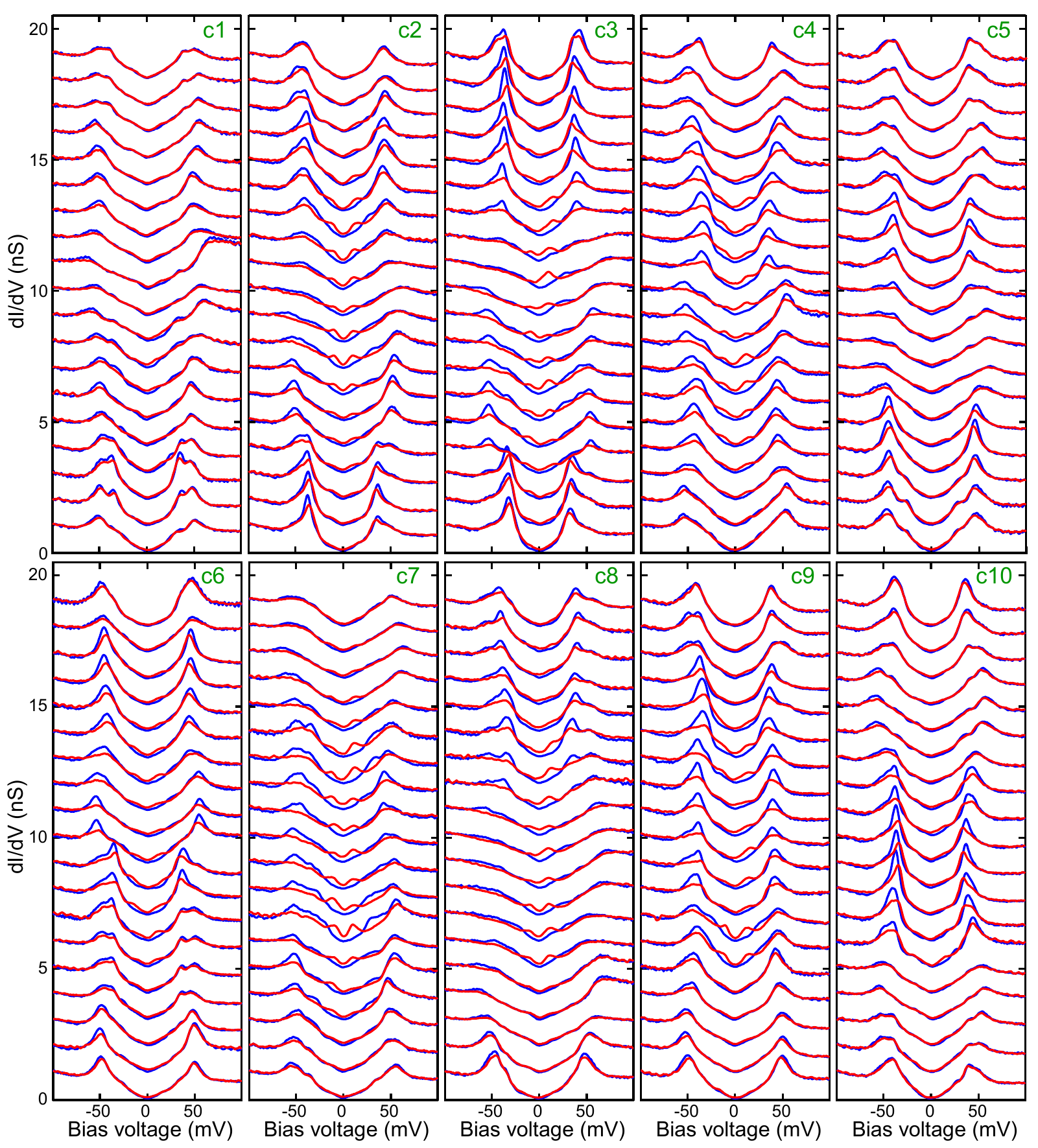}
\caption{
Line profiles of tunneling spectra along 18 paths composed of 10 paths on ``crests'' and 8 paths on ``troughs'' of the ``vortex checkerboard'' seen in $g(\mathbf{r},E=+10{\rm~meV}, B=11{\rm~T})$.
Bottoms of these line profiles correspond to the starting points of the arrows drawn in the vortex image.
Line profiles on the crests are shown in this page.
Line profiles on the troughs are shown in the previous page.
Blue and red curves in these profiles denote the tunneling spectra at $B=0$~T and 11~T, respectively.
}
\end{center}
\end{figure*}

\begin{figure*}[h]
\begin{center}
\includegraphics[width=16cm]{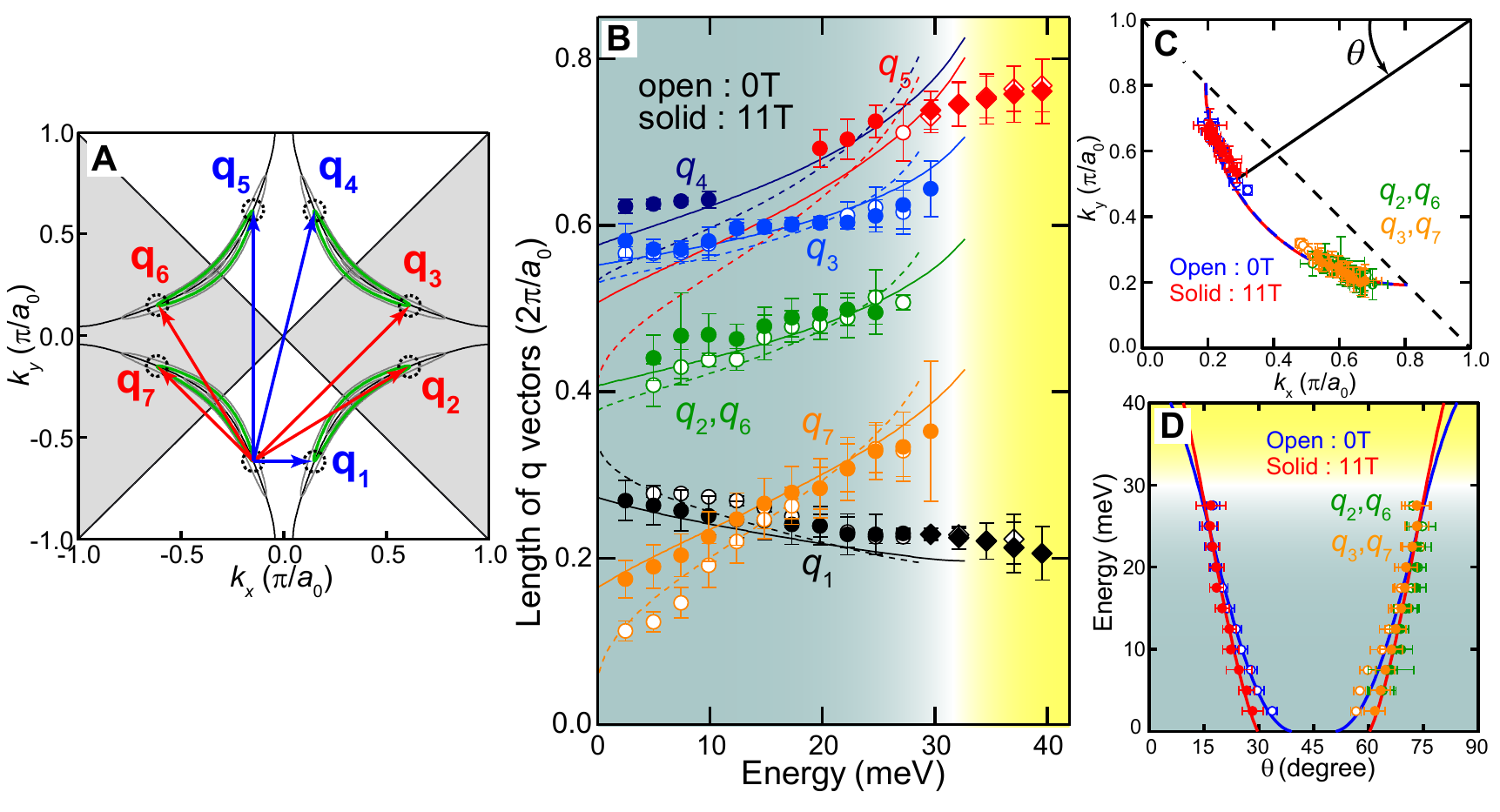}
\end{center}
\caption{
(\textbf{A}) Schematic illustration of the momentum-space electronic structure.
White and gray regions indicate the region with the opposite sign of $d$-wave superconducting gap.
Black and gray solid lines represent normal-state Fermi surface and constant energy contours of the Bogoliubov quasiparticle dispersion.
Green bold lines denote constant energy contours at a representative energy.
Black dashed circles correspond to the eight high-density-of-states regions, which primarily contribute the BQPI at a given energy.
Red and blue arrows indicate sign-reversing and sign-preserving scattering vectors~\cite{Hanaguri2009Science}.
(\textbf{B}) Energy dependence of the absolute value of the observed \textbf{q}$_{i}$'s at $B=0$~T (open symbols) and  11~T (solid symbols).
Dashed and solid lines denote the expected dispersions at $B=0$~T and 11~T, respectively.
(See Supplementary Section 2.1 for details.)
(\textbf{C}) Fermi surface loci obtained from BQPI peak locations \textbf{q}$_i$($E$), at $B=0$~T (open symbols) and 11~T (solid symbols).
Green and orange circles in right hand side of this panel are obtained from pairs of (\textbf{q}$_2$, \textbf{q}$_6$) and (\textbf{q}$_3$, \textbf{q}$_7$), respectively.
Symbols in left hand side represent the average of them.
Blue dashed and red solid lines denote the results of quarter circle fitting.
(\textbf{D}) Superconducting gap dispersion at $B=0$~T (blue symbols) and 11~T (red symbols).
Green and orange circles in right hand side of this panel are the dispersions estimated from pairs of (\textbf{q}$_2$, \textbf{q}$_6$) and (\textbf{q}$_3$, \textbf{q}$_7$), respectively.
Symbols in left hand side represent the average of them.
These are fitted to $\Delta(\theta)=\Delta_{\mathrm{BQPI}}[A\cos(2\theta)+(1-A)\cos(6\theta)]$ as shown by the solid curves.
}
\label{FigS4}
\end{figure*}

\begin{figure*}[h]
\begin{center}
\includegraphics[width=16cm]{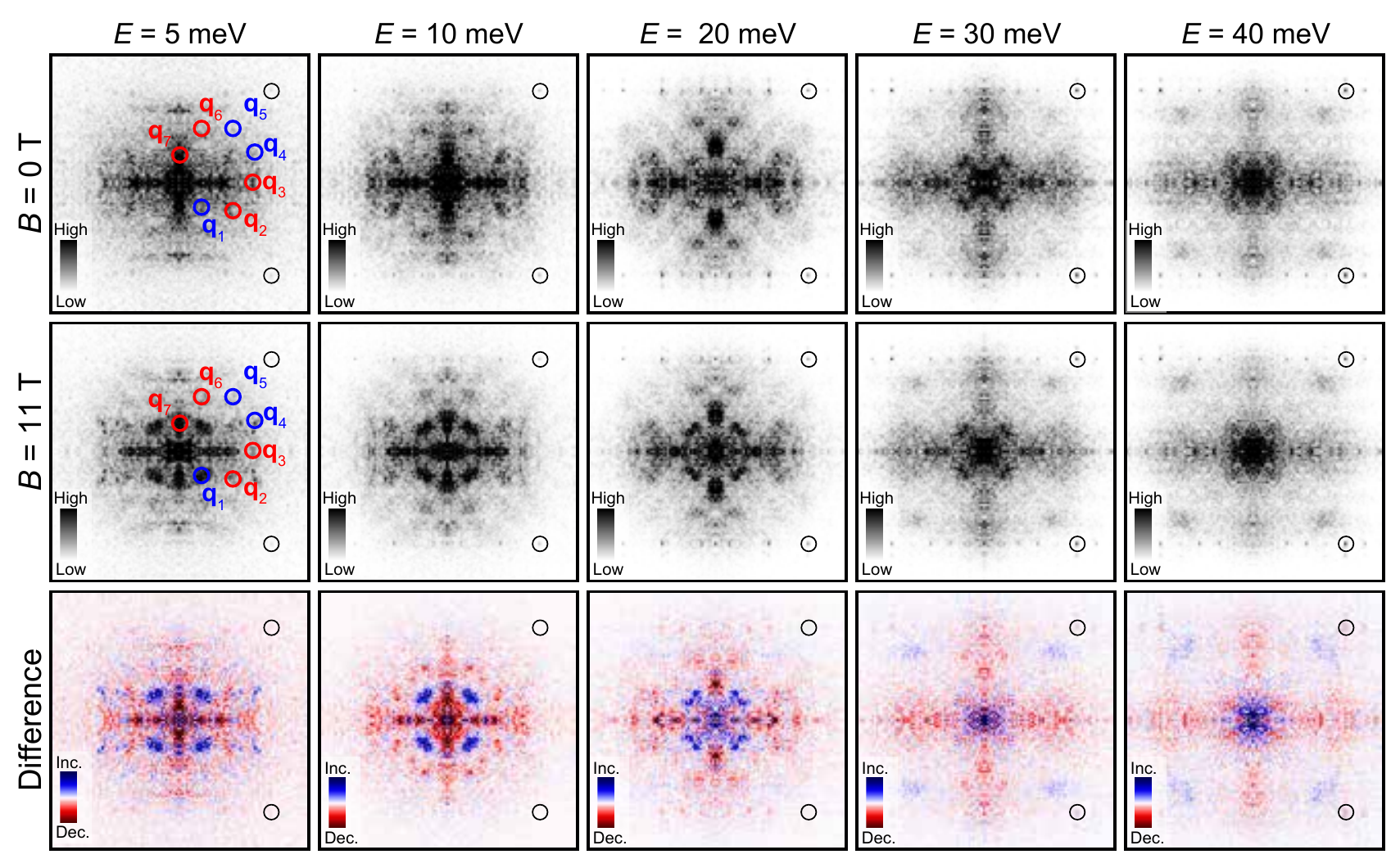}
\end{center}
\caption{
Top and middle rows display two series of $Z_q(\mathbf{q},E,B)$ at $B= 0$~T and 11~T, respectively.
Bottom row indicates difference images, which are obtained by subtracting the results at 0~T from those at 11~T.
}
\label{FigS5}
\end{figure*}

\begin{figure*}[h]
\begin{center}
\includegraphics[width=12cm]{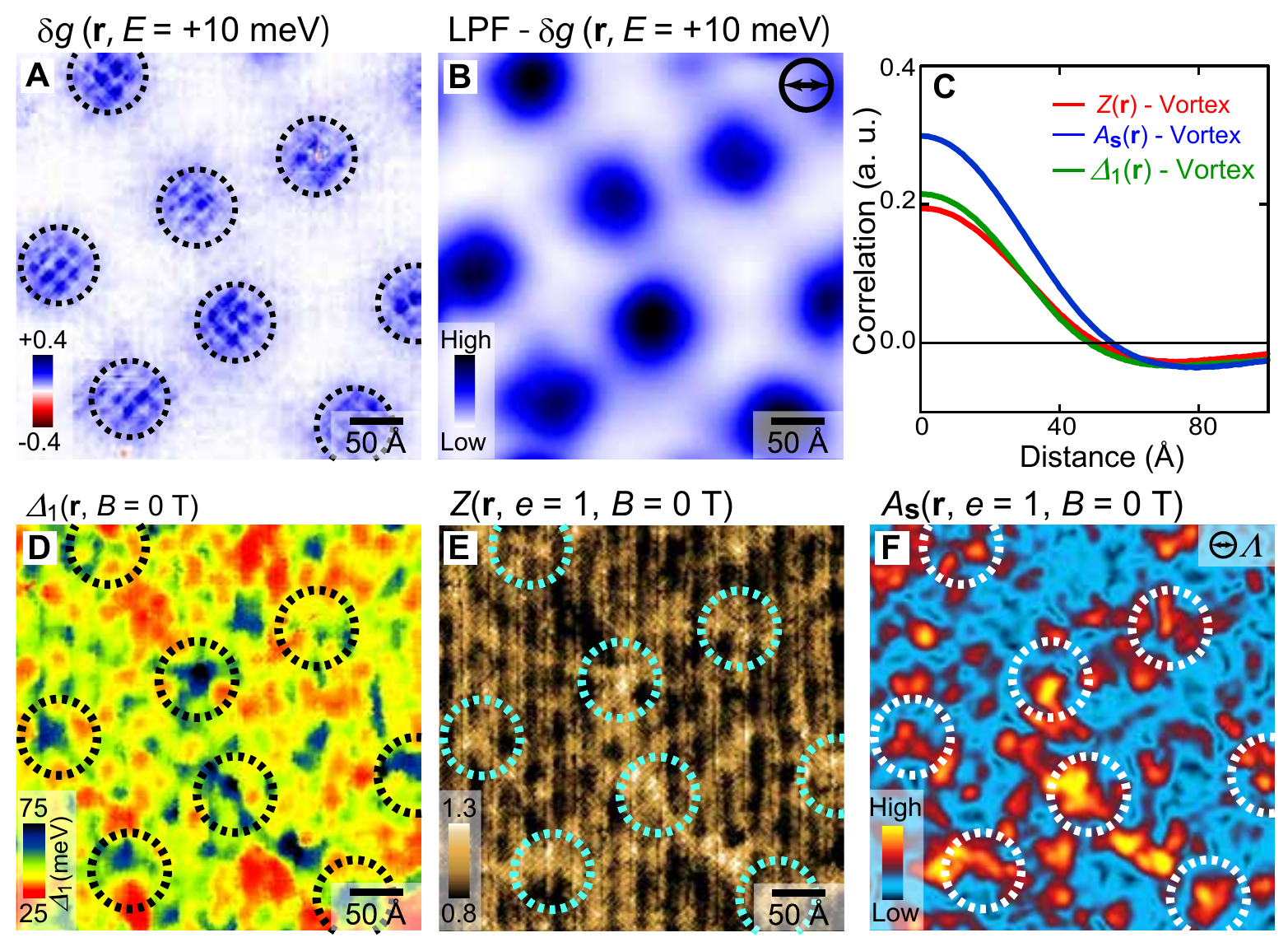}
\end{center}
\caption{
(\textbf{A}) Difference between the conductance maps at $B=0$~T and 11~T; $\delta g(\mathbf{r},E,B=11{\rm~T})=g(\mathbf{r},E,B=11{\rm~T})-g(\mathbf{r},E,B=0{\rm~T})$.
Black dashed circles mark the observed vortices.
(\textbf{B}) Low-pass filtered image of (A); LPF-$\delta g$.
A cut-off wavevector is chosen to be $0.05\times 2\pi/a_0$ in the filtering process.
Corresponding spatial resolution is indicated as a black circle and an arrow.
(\textbf{C}) Azimuthally averaged cross-correlations for pairs of [LPF-$\delta g$, $\Delta_1$] (Green curve), and [LPF-$\delta g$, $Z(\mathbf{r},e=1,B={\rm~0T})$] (Red curve), and [LPF-$\delta g$, $A_{\mathrm{S}}(\mathbf{r})$] (Blue curve).
(\textbf{D})-(\textbf{F}) $\Delta_1({\mathbf r})$, $Z(\mathbf{r},E,B=0{\rm~T})$, and $A_{\mathrm{S}}(\mathbf{r})$ at $B=0$~T.
Dashed circles in these figures represent the locations of the observed vortices.
}
\label{FigS6}
\end{figure*}

\end{document}